\documentclass[superscriptaddress,twocolumn]{revtex4-1}
\usepackage{graphicx}
\usepackage{amsmath}
\usepackage{amsfonts}
\usepackage{amssymb}
\newcommand{\be}{\begin{equation}}
\newcommand{\ee}{\end{equation}}
\newcommand{\ep}{\epsilon}
\newcommand{\R}{\mathcal{R}}

\begin{document}

\title{Full perturbative calculation of spectral correlation functions for chaotic systems in the unitary symmetry class}
\author{Sebastian M\"uller}
\affiliation{School of Mathematics, University of Bristol, Bristol BS8 1TW, UK}
\author{Marcel Novaes}
\affiliation{Instituto de F\'isica, Universidade Federal de Uberl\^andia, Uberl\^andia,
MG, 38408-100, Brazil}

\begin{abstract}

Starting from a semiclassical approach recently developed for spectral correlation
functions of quantum systems whose classical dynamics is chaotic, we focus on the case of broken time-reversal symmetry, the so-called unitary class. We obtain to all orders in
perturbation theory the non-oscillatory parts of all correlation functions, showing that the off-diagonal contributions to these correlation functions
cancel and the conjectured universality holds.
The innovation that allows this
calculation to be performed is the introduction of an auxiliary matrix model which is
governed by the same diagrammatic rules as the semiclassical approach and which can be
exactly solved.

\end{abstract}

\maketitle

\section{Introduction}

One of the central problems in the field of quantum chaos has always been to show that,
in the semiclassical limit, the energy levels of systems with chaotic classical dynamics
have local statistics that agree with the universal predictions made by random matrix
theory (RMT). This conjecture was put forward about 30 years ago \cite{bgs}, and is
supported by many numerical, experimental and theoretical results.

RMT proceeds by considering an ensemble of Hamiltonians and computing statistical
properties of the spectrum \cite{haake}. It relies only on the overall symetries of the
system, e.g. whether the Hamiltonian is real symmetric (as in systems which are
time-reversal invariant) or complex hermitian (otherwise). These ensembles of
Hamiltonians are invariant under orthogonal or unitary   transformations, respectively,
and define the   orthogonal/unitary symmetry classes. RMT is therefore a kind of
minimal-information approach and its predictions are supposed to describe `generic'
systems with no special features (in particular, having completely chaotic dynamics).

Historically, the most popular quantity to consider has been the nearest-neighbor spacing
distribution $P(s)$, but theoretically it is more convenient to work with spectral
correlation functions or their Fourier transforms. The $n$-point correlation function $\R_n(\ep_1,\cdots,\ep_n)$ measures the
likelihood that $n$ energy levels will be located around positions $E+\ep_1$, $\cdots$,
$E+\ep_n$, averaged over $E$.

The $2$-point correlation function has received most of the attention. It depends on a
single parameter ($\ep_1-\ep_2$), and a perturbation theory was developed starting from
\cite{berry}. The leading correction was derived in \cite{sieber1,sieber2}, and all
orders of perturbation theory were eventually obtained \cite{R2a,R2b,R2c}. It was even
considered beyond standard perturbation theory \cite{longa,km,longb}, recovering
oscillatory contributions to the correlation functions. Higher correlation functions
were addressed to leading order in \cite{taro}, see also \cite{shukla} for
non-oscillatory contributions.

We have made progress on this problem \cite{us} by obtaining the semiclassical
diagrammatic rules that govern the calculation of spectral correlation functions, and
showing that the simplest of these functions  indeed agree with RMT, at
least to the leading orders in perturbation theory.

In the present work we improve on this and show that, for the unitary symmetry class, the
agreement between RMT and semiclassics   persists to all correlation functions and to all orders in
perturbation theory. This is achieved through an auxiliary matrix model which is
equivalent to the semiclassical theory and which can be solved exactly.

\section{Semiclassical diagrammatic rules}

The Gutzwiller trace formula \cite{gutz} expresses the density of quantum stationary
states as a sum over classical periodic orbits. When it is used to evaluate $\R_n$, it
leads to multiple sums over periodic orbits, and the energy average selects correlations:
in order to have constructive interference one must find two sets of orbits which have
nearly the same total action.

In \cite{us} we focused attention on the case when there are $J$ orbits, denoted
$p_1,...,p_J$, correlated with another $K$ orbits, denoted $q_1,...,q_K$, so that
$J+K=n$. This leads to a kind of partial correlation function $\widetilde{\R}_{J,K}$, and
the total correlation function can be easily reconstructed from the partial ones.

The simplest possibility, called the diagonal approximation, is to have identical orbits,
i.e. the $p$ and $q$ orbits coincide pairwise, which of course is only possible for even $n$. For systems with broken
time-reversal symmetry, this was considered in \cite{taro}, where it was shown that the
diagonal approximation (in a variant that also captures oscillatory contributions)
agrees with the prediction from RMT for all $n$. That means all
corrections to this case, coming from non-identical orbits, must ultimately give a
vanishing result. This is a non-trivial fact, however, which is precisely what we want to
show.

Correlated sets of periodic orbits can be organized into diagrams. The edges of a diagram
represent long periods of time during which a $p$ orbit almost coincides with a $q$
orbit. The vertices are comparatively small regions where the orbits exchange partners
(known as `encounters' in the literature). The diagrams record only the topology of the
orbits, and their contribution to the correlation function requires integrating over all
possible action differences between the sets of orbits. This was carried out in \cite{us}
and the result is that \be\label{Rnsemi}
\widetilde{\R}_{J,K}(\epsilon,\eta)=D_{J,K}\Big[\sum_{\rm struc}\prod_{jk}
\frac{(-1)^V}{(-2\pi i(\epsilon_j-\eta_k))^{M_{jk}}}\Big].\ee
Here the energy increments included in $\epsilon=(\epsilon_1,\ldots,\epsilon_J)$
are associated to the $p$ orbits and the increments included in $\eta=(\eta_1,\ldots,\eta_K)$
are associated to the $q£$ orbits. The latter are identified with the parameters
$\epsilon_{J+1},\ldots,\epsilon_{n}$ of $\R_n$ using $\eta_k=\epsilon_{J+k}$.
Furthermore,
$M_{jk}$ is the number
of times orbits $p_j$ and $q_k$ run together but are not the first ones to arrive at an
encounter, and the derivative operator is \be\label{deriv}
D_{J,K}=\frac{(-1)^K}{(2\pi i)^n}
\prod_{j=1}^J\frac{\partial}{\partial\ep_j}\prod_{k=1}^{K}\frac{\partial}{\partial\eta_k}.\ee

The sum in (\ref{Rnsemi}) is over all the possible structures that can be associated with
a given diagram. A structure corresponds to a particular choice of order in the sequence
of traversed edges. They can be put in bijection with some equivalence classes of
factorizations of permutations, as discussed in \cite{us}. We will not follow that
approach here, but we will come back to structures in the next section.

We showed by examples in \cite{us} that, when we use the diagrammatic rule presented
above to compute $\R_n$, there are two general mechanisms leading to a vanishing result
at a particular order in perturbation theory: 1) the action of the derivative in
$D_{J,K}$ gives zero because it acts on a function with less than $n$ variables; 2)
several diagrams cancel each other. Clearly, the second mechanism is the hardest to
realize in practice. Fortunately, the model we introduce here for the unitary class
implements this mechanism automatically, leaving only the simpler possibility 1) to be
considered.

\section{Diagrammatics of a matrix model}

A direct exact evaluation of the expression in Eq. (\ref{Rnsemi}) is a hard task, because
of the complicated combinatorial problem underlying it. However, for the unitary symmetry
class we are able to make progress by following an indirect route: we postulate a matrix
model which can be treated diagrammatically with precisely the same rule.

\begin{figure}
\includegraphics[scale=0.9,clip]{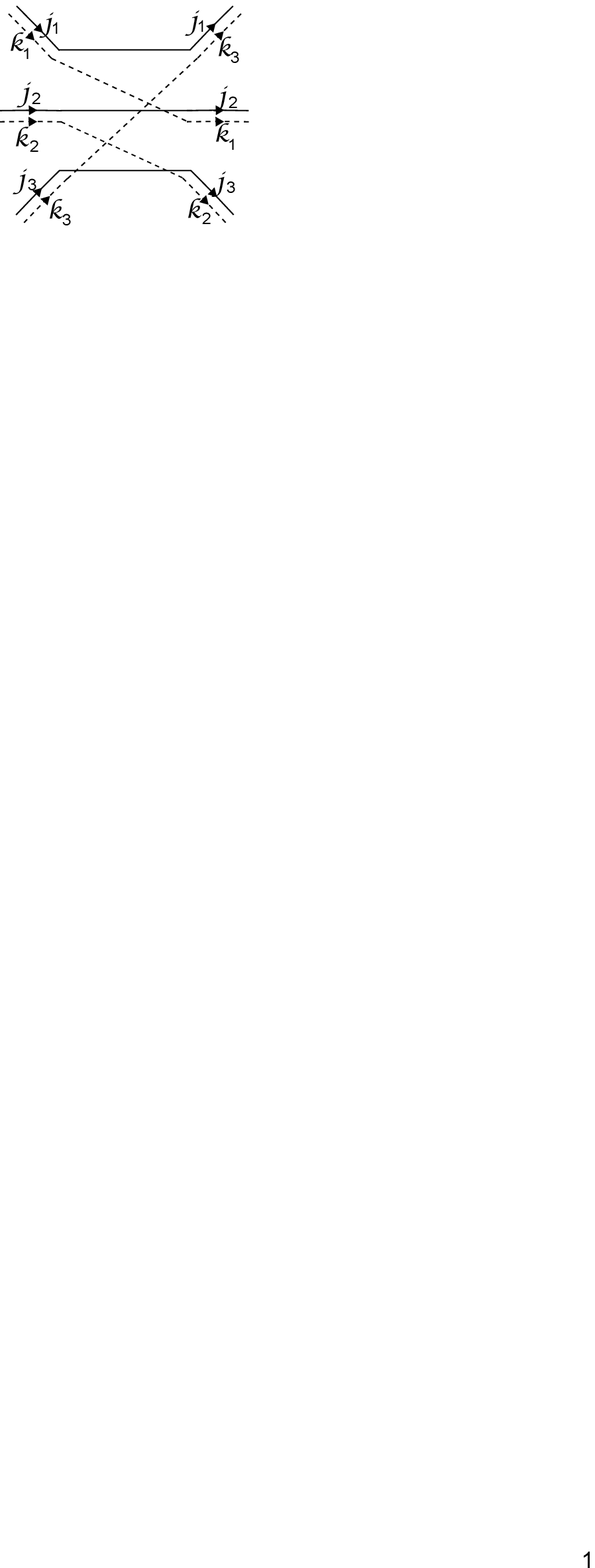}\hspace{1cm}\includegraphics[scale=0.7,clip]{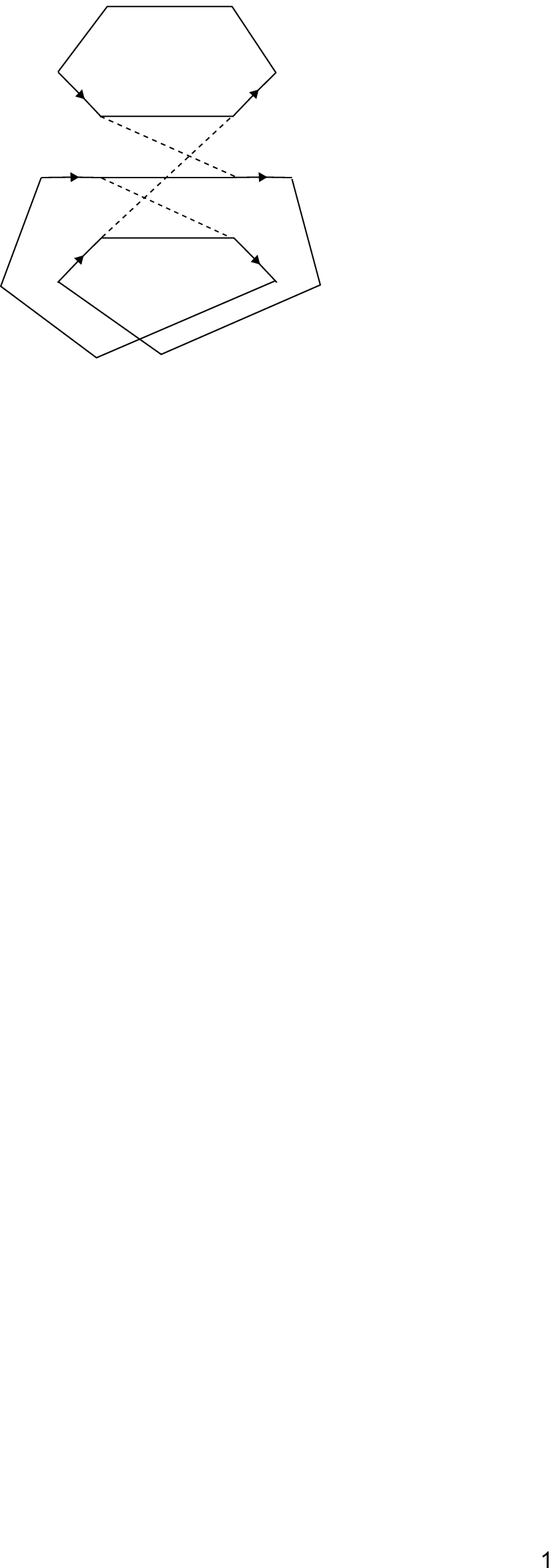}
\includegraphics[scale=0.7,clip]{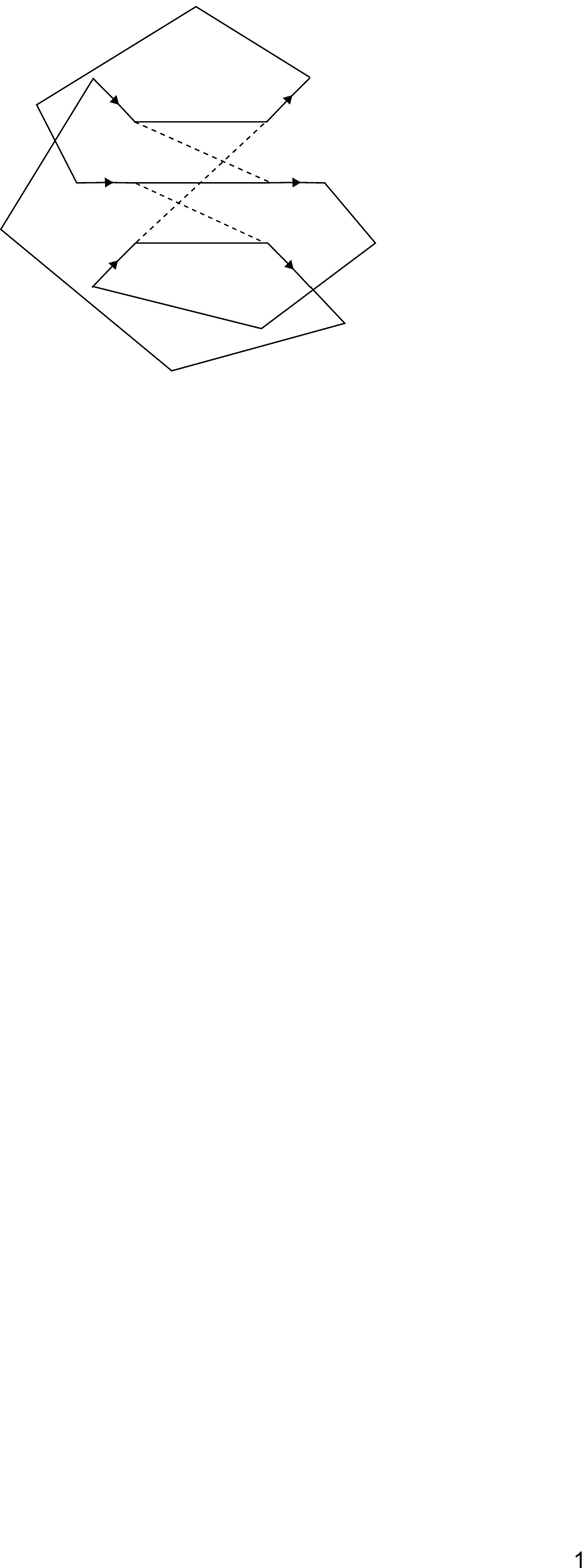}\hspace{1cm}\includegraphics[scale=0.7,clip]{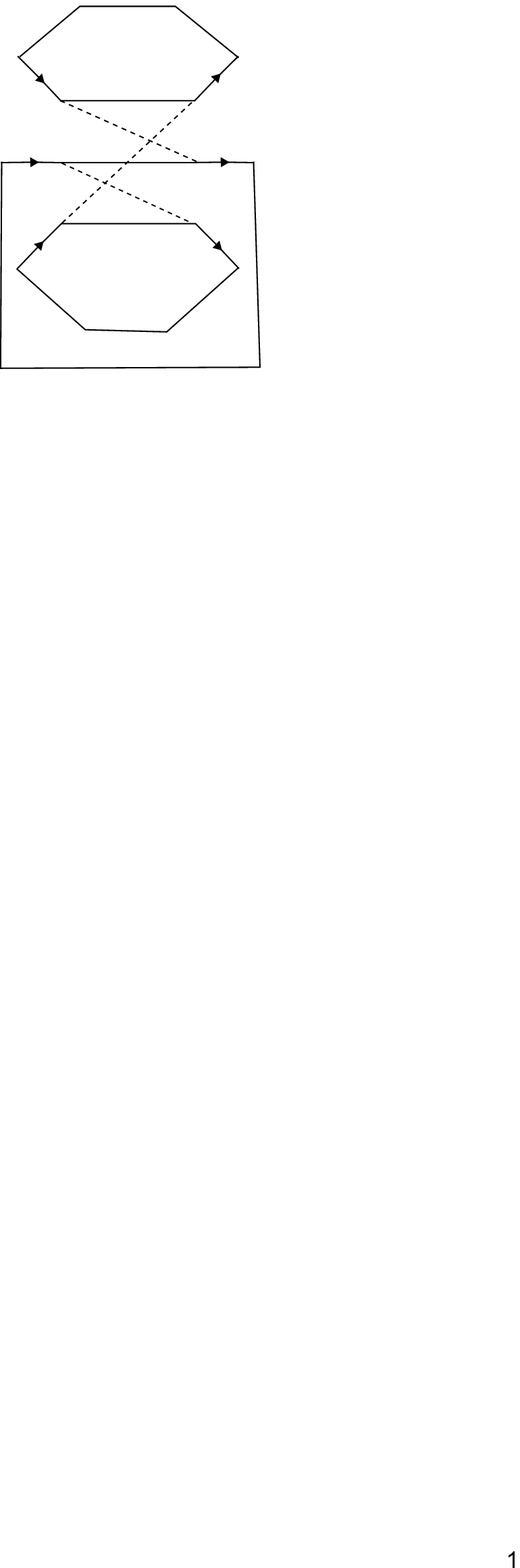}
\caption{Diagrammatics of
$Z_{j_1k_1}Z^\dag_{k_1j_2}Z_{j_2k_2}Z^\dag_{k_2j_3}Z_{j_3k_3}Z^\dag_{k_3j_1}$ (Left).
Solid lines connect incoming $j_m$ to outgoing $j_m$, while dashed lines connect incoming
$k_m$ to outgoing $k_m$. Notice how $k$'s are cyclically permuted from left to right.
Right: Sketch of a diagram contributing to $\R_4$ with $J=K=2$; notice how there are two
periodic orbits correlated with two others (encounter is grossly exaggerated).}
\end{figure}

Matrix models have a rich history, starting with the work of t' Hooft \cite{diag1}, who
noticed that Feynman diagrams in QCD could be recovered from them, and that a
perturbation theory in the inverse dimension could be arranged so as to control the
topology of these diagrams. Since then, they have been much studied
\cite{diag2a,diag2b,wittena,wittenb,wittenc}, especially in the context of 2D quantum
gravity \cite{2dqga,2dqgb,2dqgc,diag3,morozova,morozovb}.

Matrix models of a similar type to the one used here have been applied to transport
problems  \cite{transp,combinat8,energy1,energy2}. Our matrix model also has a close connection to
work relating the semiclassics of 2-point functions to the nonlinear sigma
model of RMT \cite{R2c,longb}.

We start by considering $N_1\times N_2$ complex matrices, with no
constraints -- matrix elements are independent random variables. If we
choose a Gaussian distribution \be\label{meas1} d\widetilde\mu(Z)=e^{-\alpha{\rm
Tr}(ZZ^\dag)}dZ,\ee then \be \label{cov} \frac{1}{\mathcal{Z}}\int
d\widetilde\mu(Z) Z_{jk}Z^\dag_{st}=\frac{\delta_{jt}\delta_{ks}}{\alpha},\ee where
$\mathcal{Z}=\int d\widetilde\mu(Z)$ is a normalization constant.

The Gaussian nature of the measure leads to the nice property that the average value of a
general product of matrix elements can be recovered from (\ref{cov}). This is known as
the Wick rule. It says we must sum, over all possible pairings between $Z$'s and
$Z^\dag$'s, the product of the averages of the pairs. Namely, \be \left \langle
\prod_{m=1}^q Z_{j_mk_m}Z^\dag_{s_mt_m}\right\rangle =\sum_{\pi\in
S_q}\prod_{m=1}^N\langle Z_{j_mk_m}Z^\dag_{s_{\pi(m)}t_{\pi(m)}}\rangle.\ee The sum here
is over all $q!$ permutations of the numbers from 1 to $q$.

A diagrammatic formulation can be introduced in order to compute the average value of
${\rm Tr}(ZZ^\dag)^q$. First, we expand this trace as \be
\sum_{j_1,...,j_k}\sum_{k_1,...,k_q}Z_{j_1k_1}Z^\dag_{k_1j_2}Z_{j_2k_2}Z^\dag_{k_2j_3}\cdots
Z_{j_qk_q}Z^\dag_{k_qj_1},\ee where the first sum runs from 1 to $N_1$ and the second
from $1$ to $N_2$. Then, each matrix element $Z_{jk}$ is represented as a pair of arrows,
one depicted with solid line and associated with $j$, the other depicted with dashed line
and associated with $k$. These arrows have a marked end at the tail. The matrix elements
of $Z^\dag$ are represented in the same way, but the marked end is the head. We show an
example in Figure 1. We arrange the arrows coming from $Z$ and $Z^\dag$ as two parallel
columns, and draw lines connecting coinciding indices. Finally, the diagrammatic content
of Wick's rule is that we must draw all possible connections between marked ends. The sum
over pairings becomes a sum over diagrams and the contribution of each diagram to the
average $\langle{\rm Tr}(ZZ^\dag)^q\rangle$ will result from the sum over indices of the
products of covariances like (\ref{cov}).

In the example of Figure 1 there are 6 possible ways to make the connections
allowed by Wick's rule. Three of them
are shown in the Figure. The covariances from diagram a) are such that we get the
identifications $j_2=j_3$ and $k_1=k_3$; the sum over indices gives then $N_1^2N_2^2$.
Diagram b) results in $j_1=j_2=j_3$ and $k_1=k_2=k_3$, so that its contribution is
$N_1N_2$. In diagram c) the $j$ indices have no identification, while $k_1=k_2=k_3$, so
that its contribution is $N_1^3N_2$. Considering all connections one can
show that the final result is $\langle{\rm
Tr}(ZZ^\dag)^3\rangle= \alpha^{-3}(N_1^3N_2+N_1 N_2^3+3N_1^2 N_2^2+N_1N_2)$.

When computing the average of a product of traces, each trace like ${\rm Tr}(ZZ^\dag)^q$
is represented by a vertex of valence $2q$ with a specific internal structure, such that
incoming $j_m$ is followed by outgoing $j_m$, and incoming $k_m$ is followed by outgoing
$k_m$. Wick's rule then allows connections between different vertices, producing a
diagram which may contain more than one connected component.

\section{Our matrix model}

\subsection{Definition}

The matrix integral we postulate is \be\label{integral} F(X,Y)=\frac{1}{\mathcal{Z}}\int
d\mu(Z) e^{- \sum_{q\ge 2} {\rm Tr}[X(ZZ^\dag)^q-(Z^\dag Z)^qY]},\ee where
$\mathcal{Z}=\int d\mu(Z)$ is again a suitable normalization constant. For the purposes
of this section, it is sufficient to consider $Z$ as a square matrix, of dimension $N$.
The measure $d\mu(Z)$ is still Gaussian, but given by
\be d\mu(Z)=e^{- {\rm Tr}[XZZ^\dag-Z^\dag ZY]}dZ. \ee The matrices $X$ and $Y$ are
constant and diagonal. Let us denote their eigenvalues by $x$'s and $y$'s,
respectively. The covariances of this model depend on these eigenvalues and are given by
\be \label{cov2} \frac{1}{\mathcal{Z}}\int d\mu(Z)
Z_{jk}Z^\dag_{st}=\frac{\delta_{jt}\delta_{ks}}{x_j-y_k}.\ee

\subsection{Semiclassical interpretation}

We want to give a semiclassical interpretation of our model. This will also
fix a choice for the matrices $X$ and $Y$ and a procedure to extract the
correlation functions from $F(X,Y)$. Notice that our  matrix model
only produces diagrams with encounters, i.e. corrections to the diagonal approximation of
\cite{taro}. 

The semiclassical
interpretation arises when we Taylor expand the
exponential and integrate term by term using Wick's rule to get a sum over diagrams. Then each trace/vertex can be interpreted as an encounter, e.g.
it is natural to interpret the full lines in Fig. 1 as topological representations
of how connections inside an encounter look for the $p$-orbits, and the dashed
lines as changed connections inside the $q$-orbits.

The pairwise contractions (\ref{cov2}) due to Wick's theorem become links. As discussed
previously, for a given diagram with structure, some of the $j$ indices must be
identified and likewise for the $k$ indices.
In the encounter (see Fig. 1) the $j$ indices coincide for points on the
same $p$ orbit and the $k$ indices coincide for points on the same $q$ orbit.
For a link/contraction line both indices of the connected points coincide,
in line with the fact that both must belong to the same $p$ orbit as well
as the same $q$ orbit. So it is natural to interpret the independent indices as periodic orbits, but this identification has a twist to be discussed later.

Considering the contributions, note that
the traces involve quantities of the kind $A_q=X(ZZ^\dag)^q-(Z^\dag Z)^qY.$ Notice that
\be {\rm Tr}A_q=\sum_{j_1,...,j_q}\sum_{k_1,...,k_q}(x_{j_1}-y_{k_1})
\prod_{m=1}^qZ_{j_mk_m}Z^\dag_{k_mj_{m+1}},\ee so the first indices of $X$ and $Y$ play a
distinguished role, analogous to the `first orbits to arrive at an encounter' entering the definition of $M_{jk}$ on the
semiclassical side. The denominator in (\ref{cov2}) produces a product of terms of the
form $(x_j-y_k)$ for each pair of free indices. When we take into account the similar
factor in the numerator, arising from links/contractions according to (\ref{cov2}),
we see that the contribution of a given diagram with structure
ends up being \be\label{pred} (-1)^V\prod_{jk=1}^N \frac{1}{(x_j-y_k)^{M_{jk}}}, \ee
where $M_{jk}$ is the number of times indices $j$ and $k$ belong to the same link, but
they are not the first ones to arrive at a vertex. The factor $(-1)^V$ comes from the
sign in the exponent, which produces a negative sign for each vertex.

The above diagrammatic rule looks similar to the semiclassical one (\ref{Rnsemi}).
Indeed if we could fully identify $j$ and $k$ with $p$ and $q$ orbits we
could also identify $x_j,y_k$ with the energy increments $\epsilon_j,\eta_k$
(up to a constant). This is because each orbit sum arises from the trace
formula for specific energy increments $\epsilon_j$, $\eta_k$. However
there is a crucial difference, which arises because $X$ and $Y$
are (diagonal) square matrices rather than rectangular.  In (\ref{Rnsemi}) the product over $j$ runs from 1 to $J$,
and the one over $k$ runs from 1 to $K$. On the other hand, the products in (\ref{pred})
both run from 1 to $N$. Therefore, the eigenvalues of $X$ and $Y$ cannot be directly
interpreted as energies. We must relate them to $\epsilon$'s and $\eta$'s in a
non-trivial way.

We thus extend the range of $j,k$ up to $n-1$ with $n=J+K$.
(This is the maximal range of indices we need as $j$ runs up to $J=n-1$ for
$K=1$, and $k$ runs up to $K=n-1$ for $J=1$.) 
 Moreover we include
as diagonal elements of $X$ and $Y$ several 'replicas' of these energy increments. We thus choose $N=(n-1)r$, where $n$ is the index of the correlation function we are
interested in and $r$ is some parameter. The eigenvalues of $X$ and $Y$ are then taken to
be degenerate according to \be\label{dege} x_{m+(j-1)r}=-2\pi i\epsilon_{j}, \quad
y_{m+(k-1)r}=-2\pi i\eta_{k},\ee where $m\in\{1,...,r\}$. That is, $X$ has
$n-1$ variables $\epsilon$ as eigenvalues, all $r$ times degenerate, and analogously for
$Y$ and the $\eta$ variables. When we take into account this degeneracy the diagrammatic
factor (\ref{pred}) becomes \be r^{n}(-1)^V\prod_{jk=1}^n \frac{1}{(-2\pi
i(\ep_j-\eta_k))^{M_{jk}}}.\ee

This is not exactly equal to (\ref{Rnsemi}), but it is quite close. The difference is
that here we cannot discriminate the different decompositions of $n$ as a sum $n=J+K$.
Instead, we have that if the total number of periodic orbits in a diagram is $n$, i.e. if
it must be taken into account in the calculation of $\R_n$, its contribution always gets
multiplied by $r^n$. More concretely, the function $\R_n$ can be obtained from the
function $F(X,Y)$ by first computing $[r^n]F$, the coefficient of $r^n$ in $F$. The
function $F(X,Y)$ is, by construction, a symmetric function of the $\ep$'s and a
symmetric function of the $\eta$'s. If we want, we can apply the appropriate derivative
in order to find the partial correlation function we considered before, \be
\label{tildeR}
\widetilde\R_{J,K}(\epsilon,\eta)=D_{J,K}\left([r^n]F\right).\ee

Note that on the l.h.s. $\widetilde\R_{J,K}$ should depend only on $\epsilon_j$
with $j=1,\ldots,J$ and $\eta_k$ with $k=1,\ldots,K$. However on the r.h.s.
we have not done anything to explicitly exclude contributions depending on
the remaining increments, associated to $p_j$ with $j>J$ and $q_k$ with
$k>K$. This will be justified at a later stage, when we will see that such
contributions vanish and that they do so (in a sense to be clarified then) more
immediately than $\widetilde\R_{J,K}$ itself.
     
A similar model was discussed for semiclassical chaotic transport in \cite{transp}. In
that case, all classical trajectories had the same energy. This peculiarity allowed for a
simpler matrix integral, which did not require external matrices with degenerate
eigenvalues.

\subsection{Exact solution}

Let us now proceed to the exact solution of our matrix integral $F(X,Y)$ in Eq. (\ref{integral}).

We start by computing the normalization constant. Let $Z=UDV$ be the singular value
decomposition (SVD) of $Z$. Here $D$ is a diagonal matrix such that $DD^\dag$ has the same
eigenvalues of $ZZ^\dag$, let us denote them by $\lambda$, while $U$ and $V$ are unitary
matrices. The matrix $U$ is uniformly distributed over the unitary group with Haar measure.
The matrix $V$ takes values in the coset space $U(N)/[U(1)]^N$, with a measure induced from
the Haar measure; this difference is irrelevant, leading only to a constant factor which
cancels later on.

The Jacobian of the SVD transformation was obtained in \cite{morris} and is given by
$|\Delta(\lambda)|^2$, the square of the Vandermonde determinant
$\Delta(\lambda)=\prod_{i<j}(\lambda_j-\lambda_i)$. We get \be \mathcal{Z}=\int
dUdVd\lambda |\Delta(\lambda)|^2 e^{-{\rm Tr}[XUTU^\dag]+{\rm Tr}[V^\dag TVY]},\ee where
$T=DD^\dag$. Using the well-known Harish-Chandra-Itzykson-Zuber integral
\cite{hciz1,hciz2}, \be \int dU e^{-{\rm Tr}[XUTU^\dag]}=c_N\frac{{\rm
det}(e^{-x_i\lambda_j})}{\Delta(x)\Delta(\lambda)},\ee where $c_N$ is a constant
depending only on the dimension $N$, we can now perform the integral over $U$. The integral
over $V$ is similar, but with a different constant $d_N$.

In order to do the integral over the eigenvalues, we resort to the Andr\'eief identity \cite{andre}:
given two sets of $N$ functions, $\phi_i,\psi_i$, the multidimensional integral of a
product of determinants is the determinant of a matrix whose elements are one-dimensional
integrals, \[\label{andre} \int
\det\left(\phi_j(\lambda_k)\right)\det\left(\psi_j(\lambda_k)\right)d\lambda=N!\det\int\phi_j(\lambda)\psi_k(\lambda)d\lambda.\]
Applying it to \be \int {\rm det}(e^{-x_j\lambda_k}){\rm
det}(e^{y_j\lambda_k})d\lambda\ee leads to the final result for the normalization
constant: \be \mathcal{Z}(X,Y)=\frac{c_Nd_NN!}{\Delta(x)\Delta(y)}\det C,\ee where $C$,
sometimes known as the Cauchy matrix, has elements given by $C_{jk}=(x_j-y_k)^{-1}$.

In order to compute $F(X,Y)$ we must first evaluate the sum in the exponent.
This gives
\be
F(X,Y)=\frac{1}{\mathcal{Z}}\int dUdVd\lambda |\Delta(\lambda)|^2 e^{-{\rm Tr}[XU\tilde TU^\dag]+{\rm Tr}[V^\dag \tilde TVY]}
\ee
with $\tilde T=\sum_{q\geq 1}T^q$ which has eigenvalues $\sum_{q\geq1}\lambda_i^q$.
As these diverge for $\lambda_i>1$ the integral for each $\lambda_i$ is effectively
reduced to the interval
$[0,1]$,
and the eigenvalues of the matrix $\tilde T$ become $\tilde T_i=\lambda_i/(1-\lambda_i)$
which runs from $0$ to $\infty$. The angular
integrals are computed exactly as above,
leading to
\be
F=\underbrace{\frac{1}{{\mathcal Z}}\frac{c_N d_N}{\Delta(x)\Delta(y)}}_{=\frac{1}{N!\det C}}\int\underbrace{ d\lambda\left|\frac{\Delta(\lambda)}{\Delta(\tilde T)}\right|^2}_{=d\tilde T\frac{1}{\prod_i(1+\tilde T_i)^{2N}}} \det(e^{-x_j\tilde T_K})\det(e^{y_j\tilde T_k}).
\ee
Here
 we have changed variables
to   $\tilde T_i$, using $d\lambda_i=d\tilde T_i/(1+\tilde T_i)^2$ and
\be\frac{\Delta(\lambda)}{\Delta(\tilde T)}=\prod_i\frac{1}{(1+\tilde T_i)^{N-1}}.\ee One last use of
the Andr\'eief identity (absorbing a factor $\frac{1}{(1+\tilde T)^N}$ each in $\phi_j$ and $\psi_k$) then leads to our final  result for $F(X,Y)$ as a ratio of determinants,\be
\label{Function} F(X,Y)=\frac{\det{A}}{\det C}.\ee Here the
matrix $A$ has elements given by \begin{align} A_{jk}&=
\int d\tilde T\frac{1}{(1+\tilde T)^{2N}}e^{-(x_j-y_k)\tilde T}\notag\\&=e^{x_j-y_k}{\rm Ei}(2N,x_j-y_k) ,\end{align} in terms of the incomplete exponential integral, \be {\rm Ei}(a,z) = \int_1^\infty e^{-zt}t^{-a}dt.\ee

The matrices $A$ and $C$ both have zero determinant when there are degeneracies among the $x$'s or the $y$'s. The identification between these variables and the $\epsilon$'s and $\eta$'s must therefore be performed only after computing the ratio $\det{A}/\det C$.

Integration by parts now allows us to write the result in a slightly different form,
\be\label{FCB} F(X,Y)=\frac{\det (C-B)}{\det C}\ee where \be
B_{jk}=2NC_{jk}e^{x_j-y_k}{\rm Ei}(2N+1,x_j-y_k).\ee In fact, the method of successive integration by parts can be used in order to produce an asymptotic series for each $B_{jk}$ in terms of $C_{jk}=(x_j-y_k)^{-1}$, in which higher powers of $C_{jk}$ will be accompanied by high powers of $N$.
Importantly for the following considerations, we can thus expand $F(X,Y)$ in terms of higher and higher powers of $N=(n-1)r$. The leading term of order $N^0$ is $\frac{\det C}{\det C}=1$ which vanishes after taking derivatives according to (\ref{tildeR}). All other terms involve higher powers in $N$.

\subsection{Recovering the RMT result}

Our final result (\ref{Function}) can be expanded in inverse powers of $(x-y)$ and, upon
use of the degeneracy condition (\ref{dege}), provides all spectral correlation functions. Using Eq. (\ref{tildeR}) we will show that the off-diagonal contributions to all correlation functions vanish, because the terms
that contain all variables $\epsilon_j$ and $\eta_k$ appearing in the derivatives $D_{J,K}$ must necessarily be of a higher order in $r$ than $r^n$. Hence the corresponding   coefficient $[r^n]F$ vanishes.

We start by showing this in an example. The first nontrivial term in the expansion of (\ref{FCB}) is ${\rm Tr}[C^{-1}B]$,
taken to leading order. For $n=3$ we obtain\be
Nr^2\left(\frac{1}{\ep_1-\eta_1}+\frac{1}{\ep_1-\eta_2}
+\frac{1}{\ep_2-\eta_1}+\frac{1}{\ep_2-\eta_2}\right).\ee This corresponds
semiclassically to the leading order correction to $\R_3$, as discussed in our previous
paper \cite{us}. Notice that it is indeed proportional to $r^3$ since $N=(n-1)r$. However, the contribution actually vanishes after we take the required
derivatives with respect to three different variables, since all terms inside the
parenthesis depend only on two variables.

Terms   involving all energy increments that appear in the derivatives must be of a higher order in $N$. However due to $N\propto r$ this leads to an overall order in $r$ of at least 4, and the contribution vanishes after taking the $r^3$ coefficient.

As seen in \cite{us} the second order correction to $\R_3$ involves cancellation
among several diagrams. It is not possible to see that cancellation in action in the
present matrix model: it has been performed automatically and imperceptibly.
This is precisely the merit of the model, that it has all contributions built in, and gives
only the final result.

Fortunately, we do not need to understand the function $F(X,Y)$ in its full complexity,
because of the general cancellation mechanism observed above.
In general the contributions $\widetilde \R_{J,K}$ to the $n$-point correlation function can
be accessed from $F$ using derivatives w.r.t. $n$ variables. Hence if we
expand in $(x-y)^{-1}$ the contributing terms must involve at least $n$ different
variables $x_j$, $y_k$. Because of the degeneracy the contribution is thus
proportional to $r^n$.
 However, as we have mentioned the series development of the Ei function involves further powers of $N=(n-1)r$. For all terms except the trivial leading term this leads to further factors proportional to $N$ and thus $r$, and this makes the exponent of $r$ always larger than $n$. Therefore, when taking the coefficient of order $r^n$
as required in (\ref{tildeR}), the result automatically vanishes. This is in complete agreement with the prediction from random matrix theory for the unitary symmetry class and shows that the off-diagonal contributions to all
correlation functions vanish.

To complete a technical point mentioned earlier, we can now also understand
why on the r.h.s. of Eq. (\ref{tildeR}) we did not have to explicitly exclude
contributions from orbits associated to $\epsilon_j$ with $j>J$ and $\eta_k$
with $k>K$. Any such contributions    involve one additional degeneracy
factor $r$\ for each additional energy increment. Hence these contributions
vanish even more immediately than the others and their treatment does not
affect our result.
 
 \section{Conclusions}

We have introduced a matrix model that mimics the semiclassical diagrammatic formulation
of spectral correlation functions of chaotic systems in the unitary symmetry class. This
matrix model was then solved, recovering the RMT prediction that the diagonal
approximation is exact. 

This new approach avoids cumbersome combinatorial analysis that were necessary in previous approaches to show cancellations among diagrams. Instead, these cancellations are built in the model from the start.

This is an important step forward in establishing the conjecture that spectral statistics
of quantum chaos are universal. The problem still remains to settle this conjecture for
other symmetry classes, and for the oscillatory terms which are not accessible from
perturbation theory. We hope the ideas introduced here may pave the way to the full
solution.

 SM was supported by Leverhulme Trust Research Fellowship RF-2013-470 during a part of this work. MN was supported by grants 303634/2015-4 and 400906/2016-3 from CNPq.

\end{document}